\documentclass[12pt]{article}
\usepackage{fullpage}

\title{Computation at a Distance}
\author{Samuel A.~Kutin\thanks{Center for Communications Research, 805 Bunn Drive, Princeton, NJ 08540. \hfill\break Email: {\tt \{kutin,moulton,lawren\}@idaccr.org}} \and David Petrie Moulton${}^*$ \and Lawren M.~Smithline${}^*$}
\usepackage{amsmath,amsthm,color}
\usepackage{pstricks} 
\newrgbcolor{darkgreen}{0 0.5 0}

\newdimen \myunit
\newdimen \myhsize
\newdimen \myvsize
\newcommand{\wire}[1]{\raisebox{3\myunit}[0cm][0cm]{\small #1}}
\newcommand{\smwidth}{8}
\newcommand{\smhalf}{4}
\newcommand{\qthicklines}{\linethickness{1.5\myunit}}

\newcommand{\qheight}{10}
\newcommand{\qtopheight}{5}
\newcommand{\qtopheightminusthreehalves}{3.5}
\newcommand{\qtopheightplustwo}{7}
\newcommand{\qbottomheight}{5}
\newcommand{\qbottomheightminusthreehalves}{3.5}
\newcommand{\qbottomheightplustwo}{7}

\definecolor{darkgreen}{rgb}{0,0.5,0}

\def\cn#1{
\begin{picture}(4,\qheight)(0,0)
  \put(2,\qtopheight){\circle*{2}}
  \put(0,\qtopheight){\line(1,0){4}}
\ifcase #1
    \put(2,0){\line(0,1){\qheight}}
\or \put(2,\qtopheight){\line(0,1){\qbottomheight}}
\or \put(2,\qtopheight){\line(0,-1){\qtopheight}}
\fi
\end{picture}
}

\def\ccn#1#2{
\begin{picture}(4,\qheight)(0,0)
  \put(0,\qtopheight){\line(1,0){4}}
\textcolor{#1}{
  \put(2,\qtopheight){\circle*{2}}
\ifcase #2
    \put(2,0){\line(0,1){\qheight}}
\or \put(2,\qtopheight){\line(0,1){\qbottomheight}}
\or \put(2,\qtopheight){\line(0,-1){\qtopheight}}
\fi}
\end{picture}
}

\def\nn#1{
\begin{picture}(4,\qheight)(0,0)
  \put(2,\qtopheight){\circle{3}}
  \put(0,\qtopheight){\line(1,0){.5}}
  \put(4,\qtopheight){\line(-1,0){.5}}
\ifcase #1
    \put(2,0){\line(0,1){\qtopheightminusthreehalves}}
    \put(2,\qheight){\line(0,-1){\qbottomheightminusthreehalves}}
\or \put(2,\qheight){\line(0,-1){\qbottomheightminusthreehalves}}
\or \put(2,0){\line(0,1){\qtopheightminusthreehalves}}
\fi
\end{picture}
}

\def\cnn#1#2{
\begin{picture}(4,\qheight)(0,0)
  \textcolor{#1}{\put(2,\qtopheight){\circle{3}}}
  \put(0,\qtopheight){\line(1,0){.5}}
  \put(4,\qtopheight){\line(-1,0){.5}}
\textcolor{#1}{
\ifcase #2
    \put(2,0){\line(0,1){\qtopheightminusthreehalves}}
    \put(2,\qheight){\line(0,-1){\qbottomheightminusthreehalves}}
\or \put(2,\qheight){\line(0,-1){\qbottomheightminusthreehalves}}
\or \put(2,0){\line(0,1){\qtopheightminusthreehalves}}
\fi}
\end{picture}
}

\def\xn#1{
\begin{picture}(4,\qheight)(0,0)
  \put(2,\qtopheight){\circle{4}}
  \put(0,\qtopheight){\line(1,0){4}}
\ifcase #1
    \put(2,0){\line(0,1){\qheight}}
\or \put(2,\qheight){\line(0,-1){\qbottomheightplustwo}}
\or \put(2,0){\line(0,1){\qtopheightplustwo}}
\or \put(2,\qtopheightplustwo){\line(0,-1){4}}
\fi
\end{picture}
}

\def\cxn#1#2{
\begin{picture}(4,\qheight)(0,0)
  \textcolor{#1}{
  \put(2,\qtopheight){\circle{4}}
  \put(0,\qtopheight){\line(1,0){4}}
\ifcase #2
    \put(2,0){\line(0,1){\qheight}}
\or \put(2,\qheight){\line(0,-1){\qbottomheightplustwo}}
\or \put(2,0){\line(0,1){\qtopheightplustwo}}
\or \put(2,\qtopheightplustwo){\line(0,-1){4}}
\fi}
\end{picture}
}

\newcommand{\sn}[1]{
\begin{picture}(4,\qheight)(0,0)
\ifcase #1
    \put(0,\qtopheight){\line(1,0){4}}
\or \put(2,0){\line(0,1){\qheight}}
\or \put(0,\qtopheight){\line(1,0){4}}
    \put(2,0){\line(0,1){\qheight}}
\or \put(0,\qtopheight){\line(1,0){4}}
    \multiput(2,1)(0,\qtopheight){2}{\line(0,1){\qtopheightminusthreehalves}}
\fi
\end{picture}
}

\newcommand{\csn}[2]{
\begin{picture}(4,\qheight)(0,0)
\ifcase #2
    \put(0,\qtopheight){\line(1,0){4}}
\or \textcolor{#1}{\put(2,0){\line(0,1){\qheight}}}
\or \put(0,\qtopheight){\line(1,0){4}}
    \textcolor{#1}{\put(2,0){\line(0,1){\qheight}}}
\fi
\end{picture}
}

\def\sm#1{
\begin{picture}(\smwidth,\qheight)(0,0)
\ifcase #1
    \put(0,\qtopheight){\line(1,0){\smwidth}}
\or \put(\smhalf,0){\line(0,1){\qheight}}
\or \put(0,\qtopheight){\line(1,0){\smwidth}}
    \put(\smhalf,0){\line(0,1){\qheight}}
\or \put(0,\qtopheight){\line(1,0){\smwidth}}
    \multiput(\smhalf,1)(0,\qtopheight){2}{\line(0,1){\qtopheightminusthreehalves}}
\fi
\end{picture}
}

\def\sx#1{
\begin{picture}(12,\qheight)(0,0)
\ifcase #1
    \put(0,\qtopheight){\line(1,0){12}}
\or \put(6,0){\line(0,1){\qheight}}
\or \put(0,\qtopheight){\line(1,0){12}}
    \put(6,0){\line(0,1){\qheight}}
\or \multiput(0,\qtopheight)(2.6,0){\qtopheight}{\line(1,0){1.6}}
    \put(1,0){\line(0,1){\qheight}}
    \qthicklines
    \put(11,0){\line(0,1){\qheight}}
    \thinlines
\or \multiput(0,\qtopheight)(2.6,0){\qtopheight}{\line(1,0){1.6}}
    \qthicklines
    \put(1,0){\line(0,1){\qheight}}
    \thinlines
    \put(11,0){\line(0,1){\qheight}}
\fi
\end{picture}
}

\def\dx#1{
\begin{picture}(12,\qheight)(0,0)
  \put(6,\qtopheight){\circle*{2}}
  \put(0,\qtopheight){\line(1,0){12}}
\ifcase #1
    \put(6,0){\line(0,1){\qheight}}
\or \put(6,\qtopheight){\line(0,1){\qbottomheight}}
\or \put(6,\qtopheight){\line(0,-1){\qtopheight}}
\or {}
\or \put(1,0){\line(0,1){\qheight}}
    \qthicklines
    \put(11,0){\line(0,1){\qheight}}
    \thinlines
\or \qthicklines
    \put(1,0){\line(0,1){\qheight}}
    \thinlines
    \put(11,0){\line(0,1){\qheight}}
\fi
\end{picture}
}

\def\ex#1{
\begin{picture}(12,\qheight)(0,0)
  \put(6,\qtopheight){\circle{3}}
  \put(0,\qtopheight){\line(1,0){4.5}}
  \put(12,\qtopheight){\line(-1,0){4.5}}
\ifcase #1
    \put(6,0){\line(0,1){\qtopheightminusthreehalves}}
    \put(6,\qheight){\line(0,-1){\qbottomheightminusthreehalves}}
\or \put(6,\qheight){\line(0,-1){\qbottomheightminusthreehalves}}
\or \put(6,0){\line(0,1){\qtopheightminusthreehalves}}
\or {}
\or \put(1,0){\line(0,1){\qheight}}
    \qthicklines
    \put(11,0){\line(0,1){\qheight}}
    \thinlines
\or \qthicklines
    \put(1,0){\line(0,1){\qheight}}
    \thinlines
    \put(11,0){\line(0,1){\qheight}}
\fi
\end{picture}
}

\def\nt#1{
\begin{picture}(12,\qheight)(0,0)
  \put(6,\qtopheight){\circle{4}}
  \put(0,\qtopheight){\line(1,0){12}}
\ifcase #1
    \put(6,0){\line(0,1){\qheight}}
\or \put(6,\qheight){\line(0,-1){\qbottomheightplustwo}}
\or \put(6,0){\line(0,1){\qtopheightplustwo}}
\or \put(6,\qtopheightplustwo){\line(0,-1){4}}
\or \put(1,0){\line(0,1){\qheight}}
    \qthicklines
    \put(11,0){\line(0,1){\qheight}}
    \thinlines
    \put(6,\qtopheightplustwo){\line(0,-1){4}}
\or \qthicklines
    \put(1,0){\line(0,1){\qheight}}
    \thinlines
    \put(11,0){\line(0,1){\qheight}}
    \put(6,\qtopheightplustwo){\line(0,-1){4}}
\fi
\end{picture}
}

\def\ox#1{
\begin{picture}(12,\qheight)(0,0)
  \put(6,\qtopheight){\circle{3}}
  \put(0,\qtopheight){\line(1,0){12}}
\ifcase #1 
    \put(6,0){\line(0,1){\qheight}}
\or \put(6,\qheight){\line(0,-1){6.5}}
\or \put(6,0){\line(0,1){6.5}}
\fi
\end{picture}
}

\newcommand{\ct}[1]{
\begin{picture}(12,\qheight)(0,0)
  \multiput(0,\qtopheight)(11,0){2}{\line(1,0){1}}
  \put(6,\qtopheight){\circle{\qheight}}
  \put(0,0){\vbox to \myvsize{\vfill
	\hbox to \myhsize{\hfill #1\hfill}\vfill}}
\end{picture}
}

\newcommand{\ti}[1]{
\begin{picture}(12,\qheight)(0,0)
  \multiput(1,0)(\qheight,0){2}{\line(0,1){\qheight}}
  \multiput(1,0)(0,\qheight){2}{\line(1,0){\qheight}}
  \multiput(0,\qtopheight)(11,0){2}{\line(1,0){1}}
  \put(0,0){\vbox to \myvsize{\vfill
	\hbox to \myhsize{\hfill #1\hfill}\vfill}}
\end{picture}
}

\newcommand{\tc}[1]{
\begin{picture}(12,\qheight)(0,0)
 \put(6,\qtopheight){\circle{\qheight}}
 \multiput(0,\qtopheight)(11,0){2}{\line(1,0){1}}
 \put(0,0){\vbox to \myvsize{\vfill
	\hbox to \myhsize{\hfill #1\hfill}\vfill}}
\end{picture}
}

\newcommand{\tb}[2]{
\begin{picture}(12,\qheight)(0,0)
  \put(1,0){\line(0,1){\qheight}}
  \qthicklines
  \put(11,0){\line(0,1){\qheight}}
  \thinlines
  \multiput(1,0)(\qheight,0){2}{\line(0,1){\qheight}}
  \multiput(0,\qtopheight)(11,0){2}{\line(1,0){1}}
  \put(0,0){\vbox to \myvsize{\vfill
	\hbox to \myhsize{\hfill #2\hfill}\vfill}}
\ifcase #1
{}
\or \put(1,0){\line(1,0){\qheight}}
\or \put(1,\qheight){\line(1,0){\qheight}}
\or \multiput(1,0)(0,\qheight){2}{\line(1,0){\qheight}}
\fi
\end{picture}
}

\newcommand{\tp}[2]{
\begin{picture}(12,\qheight)(0,0)
  \qthicklines
  \put(1,0){\line(0,1){\qheight}}
  \thinlines
  \put(11,0){\line(0,1){\qheight}}
  \multiput(0,\qtopheight)(11,0){2}{\line(1,0){1}}
  \put(0,0){\vbox to \myvsize{\vfill
	\hbox to \myhsize{\hfill #2\hfill}\vfill}}
\ifcase #1
{}
\or \put(1,0){\line(1,0){\qheight}}
\or \put(1,\qheight){\line(1,0){\qheight}}
\or \multiput(1,0)(0,\qheight){2}{\line(1,0){\qheight}}
\fi
\end{picture}
}

\newcommand{\place}[1]{\vbox to \myvsize{\vfill
	\hbox to \myhsize{\hfill #1\hfill}\vfill}}
\def\plac#1#2{\vbox to \myvsize{\vfill
	\hbox to #1\myhsize{#2\hfill}\vfill}}

\newcommand{\bit}[1]{\left\langle #1 \right\rangle}
\newcommand{\floor}[1]{\left\lfloor #1 \right\rfloor}
\newcommand{\ceil}[1]{\left\lceil #1 \right\rceil}
\newcommand{\xor}{\mathbin{\oplus}}
\newcommand{\xoreq}{\mathbin{\oplus\!=}}
\newcommand{\elk}{k}
\newcommand{\ff}{{\mathbf F}_2}
\newcommand{\xsum}{\bigoplus}
\newcommand{\Span}{\mathop{\mathrm{span}}}
\newcommand{\e}{{\bf e}}
\newcommand{\GL}{{\rm GL}}

\newcommand{\raiseintable}[1]{\raisebox{1.8ex}[0cm][0cm]{#1}}
\newcommand{\cnot}{{\sc cnot}}

\newtheorem{theorem}{Theorem}[section]
\newtheorem{lemma}[theorem]{Lemma}

\newtheorem{proposition}[theorem]{Proposition}
{\theoremstyle{definition}\newtheorem{definition}[theorem]{Definition}}
{\theoremstyle{remark}}

\begin{document}

\maketitle

\begin{abstract}
We consider a model of computation motivated by possible limitations
on quantum computers.  We have a linear array of $n$ wires, and we
may perform operations only on pairs of adjacent wires.  Our goal is
to build a circuits that perform specified operations spanning
all $n$ wires.  We show that the natural lower bound of $n - 1$
on circuit depth is nearly tight for a variety of problems,
and we prove linear upper bounds for additional problems.
In particular, using only gates adding a wire (mod $2$) into an
adjacent wire, we can realize any linear operation in
$\GL_n(2)$ as a circuit of depth $5n$.
We show that some linear operations require depth at least $2n+1$.
\end{abstract}

\section{Introduction}
\label{intro-sec}

We consider the following model of computation:  We have $n$ wires,
labeled $\bit{1}$ through $\bit{n}$.  Each wire carries a single bit.
We are allowed to perform reversible linear operations on adjacent
wires:\ $\bit{i} \xoreq \bit{i+1}$ or $\bit{i} \xoreq \bit{i-1}$.
We assume throughout that $n$ is at least $2$.

Our goal is to perform some calculation spanning all $n$ wires; for
example, we might want to set $\bit{n} \xoreq \bit{1}$ and leave the
other $n-2$ wires unchanged.  Our primary measure of complexity is the
{\em depth\/} of a circuit; we will also consider the {\em size\/} of
the circuit (that is, the number of gates).

The motivation for this problem is quantum circuit
design.  In some proposed models of quantum
computation~\cite{FDH,kutin,VM,VMI}, we can
perform operations only on adjacent bits, so it is important to
consider the cost of computing with bits separated by a given
distance.  Since the eventual topology of quantum computers is
unknown, we choose to focus on linear arrays of bits.  Results
here should at least be applicable to other topologies.

We note that our model is wholly classical; there are no quantum
operations.  To perform a quantum gate, one could first move bits
around using classical operations and then apply the quantum gate to
adjacent bits.  We discuss the cost of this approach in
Section~\ref{general-quantum-sec}.

It is often helpful to take an algebraic view of these circuit problems.
We adopt the convention that the wires of our circuit contain column vectors,
and we describe the state of all of the wires by the matrix whose
$i$th column is the contents of wire $\bit{i}$.  
A \cnot\ gate adds the vector on one wire into the vector on another wire.
Any circuit performs a series of column operations; note that circuits
act on the {\em right\/}.

Any function on $n$ bits that can be built out of additions may be
viewed as an element of the group $\GL_{n}(2)$ of $n\times n$
invertible matrices over the field $\ff$ of two elements.  A single gate is
represented by an elementary matrix consisting of the identity matrix
with a single $1$ either just above or just below the main diagonal.
These matrices generate the group, so we can build any reversible
linear operation on our wires using these
gates.\footnote{To implement reversible affine operations, we would
need to allow unary negation gates as well.  All such negations could be
deferred to one final time-slice.}  

It is not hard to show that any element of $\GL_{n}(2)$ can be
constructed from $O(n^2)$ gates, that
is, as a product of $O(n^2)$ of the above generators.  A simple counting
argument gives a lower bound of $\Omega(n^2/\log n)$ for generic circuits.
In Section~\ref{general-lower-sec}, we give a lower bound of $(1-o(1))n^2$
for generic elements of $\GL_{n}(2)$.

Our primary complexity measure is depth, rather than size, so the generating
set of interest is different.  We allow any set of $1$s just off the
diagonal, as long as all the row and column indices are distinct;
we cannot have
two gates using the same wire at the same time.  All of our questions
can be rephrased in this setting:  What is the shortest product of
these generators equal to a particular element of the group?

We label the wires by $\bit{1}$ through $\bit{n}$ and their
initial values by $a_1$ through $a_n$.
In our diagrams, we draw the wires horizontally,
with time proceeding from left to right, wire $\bit 1$ at the top, and
wire $\bit n$ at the bottom.  We analyze the costs of the following problems:
\begin{description}
\item[Add]  Perform $\bit{n} = a_1 \xor a_n$; for each other $i$,
leave $\bit{i} = a_i$.

\item[Swap] Set $\bit{n} = a_1$ and $\bit{1} = a_n$; for each other
  $i$, leave $\bit{i} = a_i$.

\item[Rotate] Set $\bit{n} = a_1$; for each $i < n$, set $\bit{i} = a_{i+1}$.

\item[Reverse] Set $\bit{n + 1 - i} = a_i$ for each $i$.
  
\item[Permute] Set $\bit{\sigma(i)} = a_{i}$ for each $i$, given some
  $\sigma \in S_{n}$.

\item[Compute] Apply an arbitrary $M \in \GL_n(2)$ to the $n$ wires.
\end{description}

The first two tasks require us to perform an operation on $\bit{1}$
and $\bit{n}$, leaving the other bits untouched.  The next three tasks
require us to reorder the bits; this might be useful if a quantum
circuit will perform complex calculations on different subsets of the
bits.  The final task encompasses any possible linear computation.

It is immediate that each of these tasks requires depth $n-1$, since we
need to move the information in $a_1$ at least $n-1$ times.\footnote{For
permutation and arbitrary computation, this lower bound applies in
the worst case.}
We encourage the reader to work out low-depth solutions to the above
problems before reading further.

We will prove the following results.  In each case, our proof is via
an explicit construction.

\begin{theorem}
\label{intro-add-thm}
We can add across $n$ wires in depth $n + 4$.
\end{theorem}

\begin{theorem}
\label{intro-swap-thm}
We can swap across $n$ wires in depth $n + 8$.
\end{theorem}

Our swapping circuit works by moving $a_1$ and $a_n$ to two adjacent
wires in depth roughly $n/2$, swapping the values, and then moving the
wires back.  Instead of swapping the values, we could apply any
$2$-qubit gate to the two wires.  So, we can apply any $2$-qubit quantum
gate spanning $n$ wires in depth $n + O(1)$.
In Section~\ref{general-quantum-sec}, we will generalize the above
argument.  We can apply any $m$-qubit gate whose total span is at
most $n$ in depth $n + O(m)$.

\begin{theorem}
\label{intro-rotate-thm}
We can rotate $n$ wires in depth $n + 5$.
\end{theorem}

\begin{theorem}
\label{intro-reverse-thm}
We can reverse $n$ wires in depth $2n + 2$.
\end{theorem}

We will show in Section~\ref{reverse-lower-sec} that reversal requires
a depth of at least $2n + 1$.

\begin{theorem}
\label{intro-permute-thm}
For any $\sigma \in S_{n}$, there is a circuit implementing
$\sigma$ of depth at most $3n$.
\end{theorem}


\begin{theorem}
\label{intro-gln-thm}
For any $M \in \GL_{n}(2)$, there is a circuit implementing
$M$ of depth at most~$5n$.
\end{theorem}

We will show in Section~\ref{general-lower-sec} that, for any
$\epsilon > 0$, almost every matrix in $\GL_n(2)$ requires depth
at least $(2 - \epsilon)n$.
One natural problem is to close the gap between this lower bound and
the upper bound of $5n$.  We discuss this, and other open questions,
in Section~\ref{conclude-sec}.

\suppressfloats

\section{Addition}
\label{add-sec}

\begin{theorem}
\label{add-thm}
We can add across $n$ wires in depth $n + 3$ for even $n$ and in
depth $n + 4$ for odd $n$.  The circuit has size $4n - 7$.
\end{theorem}

\begin{figure}[ht]
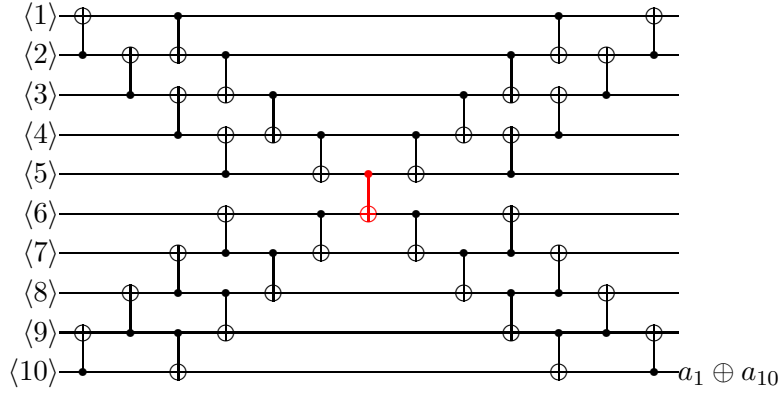

\begin{center}
\renewcommand{\arraystretch}{0}
\begin{tabular}{r@{}*{27}{c@{}}l}
\wire{$\bit{1}$} &\sn0&\xn2&\sm0&\sn0&\sm0&\cn2&\sm0&\sn0&\sm0&\sn0&\sm0&\sn0&\sm0&\sn0&\sm0&\sn0&\sm0&\sn0&\sm0&\sn0&\sm0&\cn2&\sm0&\sn0&\sm0&\xn2&\sn0& \\
\wire{$\bit{2}$} &\sn0&\cn1&\sm0&\xn2&\sm0&\xn1&\sm0&\cn2&\sm0&\sn0&\sm0&\sn0&\sm0&\sn0&\sm0&\sn0&\sm0&\sn0&\sm0&\cn2&\sm0&\xn1&\sm0&\xn2&\sm0&\cn1&\sn0& \\
\wire{$\bit{3}$} &\sn0&\sn0&\sm0&\cn1&\sm0&\xn2&\sm0&\xn1&\sm0&\cn2&\sm0&\sn0&\sm0&\sn0&\sm0&\sn0&\sm0&\cn2&\sm0&\xn1&\sm0&\xn2&\sm0&\cn1&\sm0&\sn0&\sn0& \\
\wire{$\bit{4}$} &\sn0&\sn0&\sm0&\sn0&\sm0&\cn1&\sm0&\xn2&\sm0&\xn1&\sm0&\cn2&\sm0&\sn0&\sm0&\cn2&\sm0&\xn1&\sm0&\xn2&\sm0&\cn1&\sm0&\sn0&\sm0&\sn0&\sn0& \\
\wire{$\bit{5}$} &\sn0&\sn0&\sm0&\sn0&\sm0&\sn0&\sm0&\cn1&\sm0&\sn0&\sm0&\xn1&\sm0&\ccn{red}2&\sm0&\xn1&\sm0&\sn0&\sm0&\cn1&\sm0&\sn0&\sm0&\sn0&\sm0&\sn0&\sn0& \\
\wire{$\bit{6}$} &\sn0&\sn0&\sm0&\sn0&\sm0&\sn0&\sm0&\xn2&\sm0&\sn0&\sm0&\cn2&\sm0&\cxn{red}1&\sm0&\cn2&\sm0&\sn0&\sm0&\xn2&\sm0&\sn0&\sm0&\sn0&\sm0&\sn0&\sn0& \\
\wire{$\bit{7}$} &\sn0&\sn0&\sm0&\sn0&\sm0&\xn2&\sm0&\cn1&\sm0&\cn2&\sm0&\xn1&\sm0&\sn0&\sm0&\xn1&\sm0&\cn2&\sm0&\cn1&\sm0&\xn2&\sm0&\sn0&\sm0&\sn0&\sn0& \\
\wire{$\bit{8}$} &\sn0&\sn0&\sm0&\xn2&\sm0&\cn1&\sm0&\cn2&\sm0&\xn1&\sm0&\sn0&\sm0&\sn0&\sm0&\sn0&\sm0&\xn1&\sm0&\cn2&\sm0&\cn1&\sm0&\xn2&\sm0&\sn0&\sn0& \\
\wire{$\bit{9}$} &\sn0&\xn2&\sm0&\cn1&\sm0&\cn2&\sm0&\xn1&\sm0&\sn0&\sm0&\sn0&\sm0&\sn0&\sm0&\sn0&\sm0&\sn0&\sm0&\xn1&\sm0&\cn2&\sm0&\cn1&\sm0&\xn2&\sn0& \\
\wire{$\bit{10}$} &\sn0&\cn1&\sm0&\sn0&\sm0&\xn1&\sm0&\sn0&\sm0&\sn0&\sm0&\sn0&\sm0&\sn0&\sm0&\sn0&\sm0&\sn0&\sm0&\sn0&\sm0&\xn1&\sm0&\sn0&\sm0&\cn1&\sn0& \wire{$a_1\xor{a_{10}}$} \\
\end{tabular}

\end{center}
\caption{Addition across $10$ wires ($\elk = 5$) in depth $13$.
The central \cnot\ is shown in red.}
\label{add-fig}
\end{figure}

An example of the construction for $n = 10$ appears in Figure~\ref{add-fig}.

\begin{proof}
Let $\elk = \ceil{n/2}$.  We will construct a subcircuit of depth $\elk+1$
and size $2n - 4$ that has the following effects:
\begin{enumerate}
\item $\bit{\elk} = a_1$.
\item $a_n$ contributes only to wire $\bit{\elk+1}$.
\end{enumerate}
Next, we perform $\bit{\elk+1} \xoreq \bit{\elk}$; this just replaces
$a_n$ by $a_n \xor a_1$ in the only location where $a_n$ appears.
Finally, we undo the subcircuit.  When we are done, we have
$\bit{n} = a_n \xor a_1$, and each other wire has its initial value.
The overall circuit size is $4n - 7$, and the depth is $2\elk + 3$ as
desired.

\begin{figure}
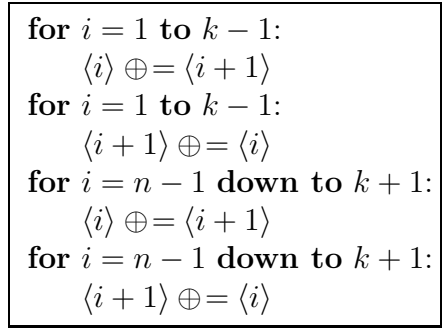

\centerline{\fbox{
\begin{minipage}{100in}
\begin{tabbing}
{\bf for} \= {\bf for} \= {\bf for} \= \kill
{\bf for} $i = 1$ {\bf to} $\elk - 1$: \\
\> $\bit{i} \xoreq \bit{i+1}$ \\
{\bf for} $i = 1$ {\bf to} $\elk - 1$: \\
\> $\bit{i+1} \xoreq \bit{i}$ \\
{\bf for} $i = n-1$ {\bf down to} $\elk+1$:  \\
\> $\bit{i} \xoreq \bit{i+1}$ \\
{\bf for} $i = n-1$ {\bf down to} $\elk+1$:  \\
\> $\bit{i+1} \xoreq \bit{i}$
\end{tabbing}
\end{minipage}}}
\caption{Subcircuit for addition.  We take $\elk = \ceil{n/2}$.}
\label{add-code-fig}
\end{figure}

It remains only to discuss the subcircuit, which is described in
Figure~\ref{add-code-fig}.  We begin with the first two loops, or
``cascades''.  The first loop writes $a_i \xor a_{i+1}$ to $\bit{i}$
for $i < \elk$.  After the second loop, $\bit{i}$ contains $a_1 \xor
a_{i+1}$ for $i < \elk$, and $\bit{\elk}$ contains $a_1$.  Notice that
we can start the second loop during the third time-slice, so the two
cascades together have depth $\elk + 1$.

The third and fourth loops can be similarly analyzed.  After both
loops are completed, we have written $a_{i-1}$ to $\bit{i}$ (for $i >
\elk + 1$) and $\xsum_{j=\elk+1}^n a_j$ to $\bit{\elk+1}$.  As
desired, $a_n$ affects only $\bit{\elk+1}$.  The depth is $(n - 1 -
\elk) + 2 \le \elk + 1$.
\end{proof}

\suppressfloats

\section{Swap}
\label{swap-sec}

\begin{theorem}
\label{swap-thm}
We can swap across $n$ wires in depth $n + 7$ for even $n$ and
in depth $n+8$ for odd $n$.  The circuit has size $6n - 9$.
\end{theorem}

\begin{figure}
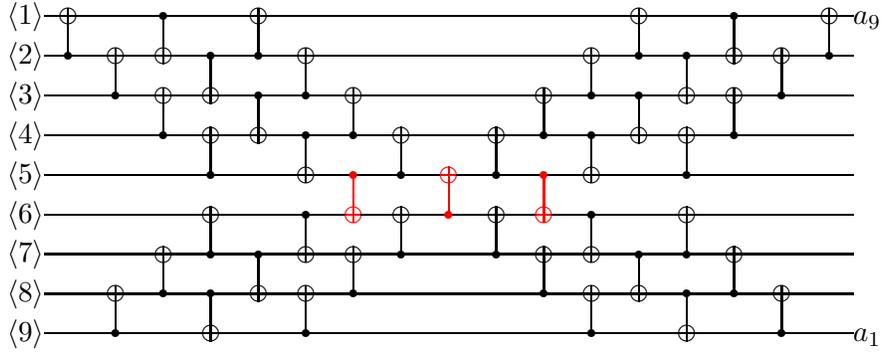

\begin{center}
\renewcommand{\arraystretch}{0}
\begin{tabular}{r@{}*{35}{c@{}}l}
\wire{$\bit{1}$} &\sn0&\xn2&\sm0&\sn0&\sm0&\cn2&\sm0&\sn0&\sm0&\xn2&\sm0&\sn0&\sm0&\sn0&\sm0&\sn0&\sm0&\sn0&\sm0&\sn0&\sm0&\sn0&\sm0&\sn0&\sm0&\xn2&\sm0&\sn0&\sm0&\cn2&\sm0&\sn0&\sm0&\xn2&\sn0& \wire{$a_9$} \\
\wire{$\bit{2}$} &\sn0&\cn1&\sm0&\xn2&\sm0&\xn1&\sm0&\cn2&\sm0&\cn1&\sm0&\xn2&\sm0&\sn0&\sm0&\sn0&\sm0&\sn0&\sm0&\sn0&\sm0&\sn0&\sm0&\xn2&\sm0&\cn1&\sm0&\cn2&\sm0&\xn1&\sm0&\xn2&\sm0&\cn1&\sn0& \\
\wire{$\bit{3}$} &\sn0&\sn0&\sm0&\cn1&\sm0&\xn2&\sm0&\xn1&\sm0&\cn2&\sm0&\cn1&\sm0&\xn2&\sm0&\sn0&\sm0&\sn0&\sm0&\sn0&\sm0&\xn2&\sm0&\cn1&\sm0&\cn2&\sm0&\xn1&\sm0&\xn2&\sm0&\cn1&\sm0&\sn0&\sn0& \\
\wire{$\bit{4}$} &\sn0&\sn0&\sm0&\sn0&\sm0&\cn1&\sm0&\xn2&\sm0&\xn1&\sm0&\cn2&\sm0&\cn1&\sm0&\xn2&\sm0&\sn0&\sm0&\xn2&\sm0&\cn1&\sm0&\cn2&\sm0&\xn1&\sm0&\xn2&\sm0&\cn1&\sm0&\sn0&\sm0&\sn0&\sn0& \\
\wire{$\bit{5}$} &\sn0&\sn0&\sm0&\sn0&\sm0&\sn0&\sm0&\cn1&\sm0&\sn0&\sm0&\xn1&\sm0&\ccn{red}2&\sm0&\cn1&\sm0&\cxn{red}2&\sm0&\cn1&\sm0&\ccn{red}2&\sm0&\xn1&\sm0&\sn0&\sm0&\cn1&\sm0&\sn0&\sm0&\sn0&\sm0&\sn0&\sn0& \\
\wire{$\bit{6}$} &\sn0&\sn0&\sm0&\sn0&\sm0&\sn0&\sm0&\xn2&\sm0&\sn0&\sm0&\cn2&\sm0&\cxn{red}1&\sm0&\xn2&\sm0&\ccn{red}1&\sm0&\xn2&\sm0&\cxn{red}1&\sm0&\cn2&\sm0&\sn0&\sm0&\xn2&\sm0&\sn0&\sm0&\sn0&\sm0&\sn0&\sn0& \\
\wire{$\bit{7}$} &\sn0&\sn0&\sm0&\sn0&\sm0&\xn2&\sm0&\cn1&\sm0&\cn2&\sm0&\xn1&\sm0&\xn2&\sm0&\cn1&\sm0&\sn0&\sm0&\cn1&\sm0&\xn2&\sm0&\xn1&\sm0&\cn2&\sm0&\cn1&\sm0&\xn2&\sm0&\sn0&\sm0&\sn0&\sn0& \\
\wire{$\bit{8}$} &\sn0&\sn0&\sm0&\xn2&\sm0&\cn1&\sm0&\cn2&\sm0&\xn1&\sm0&\xn2&\sm0&\cn1&\sm0&\sn0&\sm0&\sn0&\sm0&\sn0&\sm0&\cn1&\sm0&\xn2&\sm0&\xn1&\sm0&\cn2&\sm0&\cn1&\sm0&\xn2&\sm0&\sn0&\sn0& \\
\wire{$\bit{9}$} &\sn0&\sn0&\sm0&\cn1&\sm0&\sn0&\sm0&\xn1&\sm0&\sn0&\sm0&\cn1&\sm0&\sn0&\sm0&\sn0&\sm0&\sn0&\sm0&\sn0&\sm0&\sn0&\sm0&\cn1&\sm0&\sn0&\sm0&\xn1&\sm0&\sn0&\sm0&\cn1&\sm0&\sn0&\sn0& \wire{$a_1$} \\
\end{tabular}

\end{center}
\caption{Swap across $9$ wires ($\elk = 5$) in depth $17$.
The central swap is shown in red.} 
\label{swap-fig}
\end{figure}

An example of this construction for $n = 9$ appears in
Figure~\ref{swap-fig}.

\begin{proof}
We use the same basic idea as in the proof of Theorem~\ref{add-thm}.
As before, let $\elk = \ceil{n/2}$.  Before, we built a subcircuit
guaranteeing that $\bit{\elk} = a_1$ and that $a_n$ contributes only
to wire $\bit{\elk+1}$.  For a swap, we need something stronger:
\begin{enumerate}
\item $\bit{\elk} = a_1$.
\item $\bit{\elk+1} = a_n$.
\item No other wire depends on $a_1$ or $a_n$.
\end{enumerate}
Our subcircuit will have size $3n - 6$ and depth $\elk + 3$.

We begin by running the subcircuit.  Next, we swap $\bit{\elk}$ and
$\bit{\elk+1}$; this requires three gates.  Finally, we undo the subcircuit.
The overall size is $6n - 9$.

\begin{figure}
\centerline{\fbox{
\begin{minipage}{100in}
\begin{tabbing}
{\bf for} \= {\bf for} \= {\bf for} \= \kill
{\bf for} $i = 1$ {\bf to} $\elk-1$: \\
\> $\bit{i} \xoreq \bit{i+1}$ \\
{\bf for} $i = 1$ {\bf to} $\elk-1$: \\
\> $\bit{i+1} \xoreq \bit{i}$ \\
{\bf for} $i = 1$ {\bf to} $\elk-1$: \\
\> $\bit{i} \xoreq \bit{i+1}$ \\
{\bf for} $i = n-1$ {\bf down to} $\elk+1$:  \\
\> $\bit{i} \xoreq \bit{i+1}$ \\
{\bf for} $i = n-1$ {\bf down to} $\elk+1$:  \\
\> $\bit{i+1} \xoreq \bit{i}$ \\
{\bf for} $i = n-1$ {\bf down to} $\elk+1$:  \\
\> $\bit{i} \xoreq \bit{i+1}$
\end{tabbing}
\end{minipage}}}
\caption{Subcircuit for swap.  We take $\elk = \ceil{n/2}$.}
\label{swap-code-fig}
\end{figure}

The subcircuit is described in Figure~\ref{swap-code-fig}.  The first
two loops are the same as in Figure~\ref{add-code-fig}.  We write $a_1
+ a_{i+1}$ to $\bit{i}$ (for $i < \elk$) and $a_1$ to $\bit{\elk}$.
The next loop erases the $a_1$ information; when it concludes, we have
$\bit{i} = a_{i+1} + a_{i+2}$ for $i < \elk - 1$, $\bit{\elk-1} =
a_{\elk}$, and $\bit{\elk} = a_1$.  As before, we can nest the cascades
(see Figure~\ref{swap-fig}); the depth is~$\elk+3$.

The remaining loops are similar.  After the penultimate loop, we have
$\bit{i} = a_{i-1}$ for $i > \elk+1$ and $\bit{\elk+1} =
\xsum_{j=\elk+1}^n a_j$.  The final loop accumulates upward; we obtain
$\bit{i} = \xsum_{j=i-1}^{n-1} a_j$ for $i > \elk+1$, and
$\bit{\elk+1} = a_n$.  The depth is $(n - 1 - \elk) + 4 \le \elk + 3$.

Since the subcircuit has depth $\elk + 3$, and the central swap has
depth $3$, one might think the overall depth would be $2\elk + 9$.
In fact, we can reduce the depth to $2\elk + 7$.  Two of the three
gates in the swap commute with adjacent gates and can be nested into
the subcircuit, as shown in Figure~\ref{swap-fig}.
\end{proof}

\subsection{Arbitrary Quantum Gates}
\label{general-quantum-sec}

As noted in the Introduction, we could replace the central swap with
any operation on $a_1$ and $a_n$; in the quantum setting, we could use
any $2$-qubit gate.  Hence, any $2$-qubit gate spanning $n$ wires can
be implemented in depth~$n + O(1)$.

Suppose that we wish to implement an $m$-qubit gate with span $n$.  We
need to operate on a set of bits $\bit{i_1}, \dots, \bit{i_m}$
with $1 = i_1 < i_2 < \cdots < i_m = n$.  Write $b_\ell = a_{i_\ell}$.
Let $k = \ceil{n/2}$ as above, and choose $j$ with $i_j \le k < i_{j+1}$.

For each $\ell$ between $1$ and $m$, we will move $b_\ell$ onto the
wire $\bit{k - j + \ell}$, so the bits will lie on $m$ adjacent wires.
We then perform the $m$-qubit gate. Finally, we undo the
transformation.

We will begin with nested cascades as in our swap circuit; we use the
top half of the subcircuit of Figure~\ref{swap-code-fig}, but we only
let $i$ range from $i_j$ to $k-1$.  When we finish, we have $\bit{k} =
b_j$, and no other wire depends on $b_j$.  The wires between $\bit{j}$
and $\bit{k-1}$ contain some complicated functions of various $a_i$ bits,
but none of the $b_\ell$ bits are involved.

Next, if $j > 1$, we perform cascades moving $b_{j-1}$ to $\bit{k-1}$.
We continue, performing a series of $j$ sets of cascades; the final
set moves $b_1$ into $\bit{k - j + 1}$.  Since the cascades nest, the
total depth is~$k + O(m)$.

At the same time, we perform upward cascades moving $b_{j+1}$ to
$\bit{k+1}$, $b_{j+2}$ to $\bit{k+2}$, and so on, up to moving $b_m$
to $\bit{k - j + m}$.  After $k + O(m)$ time-slices, we have moved the
$m$ bits of interest onto the wires from $\bit{k-j+1}$
to~$\bit{k-j+m}$.

Finally, we perform the $m$-qubit gate, and we reverse the first part
of the computation to put all the bits back.  The overall
depth is $n + O(m)$, in addition to the cost of the $m$-qubit quantum gate.

Moreover, suppose we wish to perform several long-range gates spanning
$n$ wires, and using a total of $m$ bits, simultaneously.  We first
move those $m$ bits together in depth $n + O(m)$.  Next, we permute
the bits in depth $O(m)$ (see Section~\ref{permute-sec}), so the bits
for each gate are adjacent.  We now perform the quantum gates and then
undo the rest of the calculation.  The total depth is again $n + O(m)$,
in addition to the cost of the most complicated quantum gate.

\section{Rotation}
\label{rotate-sec}

Recall that rotating $n$ wires means setting $\bit{n}$ to $a_1$
and setting $\bit{i}$ to $a_{i+1}$ for each other $i$.

\begin{theorem}
\label{rotate-thm}
For $n > 2$, we can rotate $n$ wires in depth $n + 5$.
The circuit has size $4n - 6$.
\end{theorem}

\begin{figure}[hb]
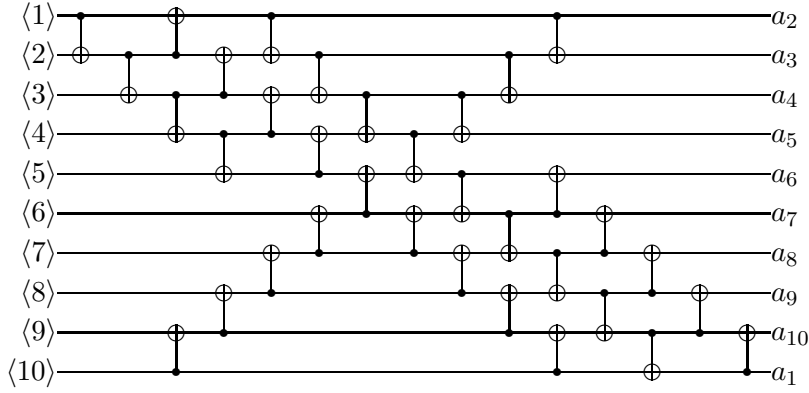

\begin{center}
\renewcommand{\arraystretch}{0}
\begin{tabular}{r@{}*{31}{c@{}}l}
\wire{$\bit{1}$} &\sn0&\cn2&\sm0&\sn0&\sm0&\xn2&\sm0&\sn0&\sm0&\cn2&\sm0&\sn0&\sm0&\sn0&\sm0&\sn0&\sm0&\sn0&\sm0&\sn0&\sm0&\cn2&\sm0&\sn0&\sm0&\sn0&\sm0&\sn0&\sm0&\sn0&\sn0& \wire{$a_2$} \\
\wire{$\bit{2}$} &\sn0&\xn1&\sm0&\cn2&\sm0&\cn1&\sm0&\xn2&\sm0&\xn1&\sm0&\cn2&\sm0&\sn0&\sm0&\sn0&\sm0&\sn0&\sm0&\cn2&\sm0&\xn1&\sm0&\sn0&\sm0&\sn0&\sm0&\sn0&\sm0&\sn0&\sn0& \wire{$a_3$} \\
\wire{$\bit{3}$} &\sn0&\sn0&\sm0&\xn1&\sm0&\cn2&\sm0&\cn1&\sm0&\xn2&\sm0&\xn1&\sm0&\cn2&\sm0&\sn0&\sm0&\cn2&\sm0&\xn1&\sm0&\sn0&\sm0&\sn0&\sm0&\sn0&\sm0&\sn0&\sm0&\sn0&\sn0& \wire{$a_4$} \\
\wire{$\bit{4}$} &\sn0&\sn0&\sm0&\sn0&\sm0&\xn1&\sm0&\cn2&\sm0&\cn1&\sm0&\xn2&\sm0&\xn1&\sm0&\cn2&\sm0&\xn1&\sm0&\sn0&\sm0&\sn0&\sm0&\sn0&\sm0&\sn0&\sm0&\sn0&\sm0&\sn0&\sn0& \wire{$a_5$} \\
\wire{$\bit{5}$} &\sn0&\sn0&\sm0&\sn0&\sm0&\sn0&\sm0&\xn1&\sm0&\sn0&\sm0&\cn1&\sm0&\xn2&\sm0&\xn1&\sm0&\cn2&\sm0&\sn0&\sm0&\xn2&\sm0&\sn0&\sm0&\sn0&\sm0&\sn0&\sm0&\sn0&\sn0& \wire{$a_6$} \\
\wire{$\bit{6}$} &\sn0&\sn0&\sm0&\sn0&\sm0&\sn0&\sm0&\sn0&\sm0&\sn0&\sm0&\xn2&\sm0&\cn1&\sm0&\xn2&\sm0&\xn1&\sm0&\cn2&\sm0&\cn1&\sm0&\xn2&\sm0&\sn0&\sm0&\sn0&\sm0&\sn0&\sn0& \wire{$a_7$} \\
\wire{$\bit{7}$} &\sn0&\sn0&\sm0&\sn0&\sm0&\sn0&\sm0&\sn0&\sm0&\xn2&\sm0&\cn1&\sm0&\sn0&\sm0&\cn1&\sm0&\xn2&\sm0&\xn1&\sm0&\cn2&\sm0&\cn1&\sm0&\xn2&\sm0&\sn0&\sm0&\sn0&\sn0& \wire{$a_8$} \\
\wire{$\bit{8}$} &\sn0&\sn0&\sm0&\sn0&\sm0&\sn0&\sm0&\xn2&\sm0&\cn1&\sm0&\sn0&\sm0&\sn0&\sm0&\sn0&\sm0&\cn1&\sm0&\xn2&\sm0&\xn1&\sm0&\cn2&\sm0&\cn1&\sm0&\xn2&\sm0&\sn0&\sn0& \wire{$a_9$} \\
\wire{$\bit{9}$} &\sn0&\sn0&\sm0&\sn0&\sm0&\xn2&\sm0&\cn1&\sm0&\sn0&\sm0&\sn0&\sm0&\sn0&\sm0&\sn0&\sm0&\sn0&\sm0&\cn1&\sm0&\xn2&\sm0&\xn1&\sm0&\cn2&\sm0&\cn1&\sm0&\xn2&\sn0& \wire{$a_{10}$} \\
\wire{$\bit{10}$} &\sn0&\sn0&\sm0&\sn0&\sm0&\cn1&\sm0&\sn0&\sm0&\sn0&\sm0&\sn0&\sm0&\sn0&\sm0&\sn0&\sm0&\sn0&\sm0&\sn0&\sm0&\cn1&\sm0&\sn0&\sm0&\xn1&\sm0&\sn0&\sm0&\cn1&\sn0& \wire{$a_1$} \\
\end{tabular}

\end{center}
\caption{Rotation of $10$ wires ($\elk = 5$) in depth $15$.}
\label{rotate-fig}
\end{figure}

We first give a rotation circuit of depth $2n + 1$.  We then
explain how to use this circuit in our main construction.
An example of the final result with $n = 10$ is depicted in
Figure~\ref{rotate-fig}.

\begin{lemma}
\label{rotate-2d-lemma}
We can rotate $n$ wires in depth $2n + 1$.  The circuit has
size~$4n - 5$.
\end{lemma}

\begin{proof}
We consider the rotation circuit of Figure~\ref{rotate-code-fig},
which we call~$R(\ell, m)$.

\begin{figure}
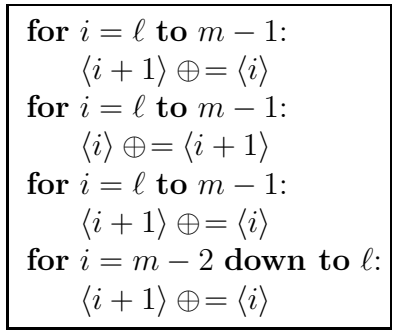

\centerline{\fbox{
\begin{minipage}{100in}
\begin{tabbing}
{\bf for} \= {\bf for} \= {\bf for} \= \kill
{\bf for} $i = \ell$ {\bf to} $m-1$: \\
\> $\bit{i+1} \xoreq \bit{i}$ \\
{\bf for} $i = \ell$ {\bf to} $m-1$: \\
\> $\bit{i} \xoreq \bit{i+1}$ \\
{\bf for} $i = \ell$ {\bf to} $m-1$: \\
\> $\bit{i+1} \xoreq \bit{i}$ \\
{\bf for} $i = m-2$ {\bf down to} $\ell$: \\
\> $\bit{i+1} \xoreq \bit{i}$
\end{tabbing}
\end{minipage}}}
\caption{Rotation circuit $R(\ell,m)$.}
\label{rotate-code-fig}
\end{figure}

The first three loops of $R(\ell,m)$ are similar to those in
Figure~\ref{swap-code-fig}.  After the first loop, we have
$\bit{j} = \xsum_{i=\ell}^j a_i$ for $\ell \le j \le m$.  The
second loop leaves $\bit{m} = \xsum_{i=\ell}^m a_i$, and sets each
other $\bit{j}$ to $a_{j+1}$.  The third loop sets $\bit{j}$ to
$\xsum_{i=\ell+1}^{j+1} a_j$ for $j < m$, but sets $\bit{m} = a_\ell$.
The final loop restores $\bit{j}$ to $a_{j+1}$ for $j < m$.

The circuit $R(\ell, m)$ contains $4(m-\ell) - 1$ gates.  The first
three loops can be nested, for a combined depth of $(m-\ell) + 4$.
The total depth is $2(m-\ell) + 3$.  If we take $\ell = 1$ and $m =
n$, we obtain a rotation of all $n$ wires.
\end{proof}

Note that if we flip each gate in $R(\ell, m)$
upside-down, the resulting circuit still performs a rotation.
More generally, the circuit formed by flipping each gate of a given circuit
upside-down performs the inverse transpose of the $\GL_n(2)$ transformation
performed by the original circuit.

\begin{proof}[Proof of Theorem~\ref{rotate-thm}]
Let $k = \ceil{n/2}$.  We let $R(\ell, m)$ be the circuit of 
Lemma~\ref{rotate-2d-lemma}.

We let $R'(\ell, m)$ be the circuit $R(\ell, m)$ run upside-down and
backward.  Note that running a rotation circuit upside-down makes it
rotate in the opposite direction, and running any circuit backward
makes it perform the inverse operation.  So, $R'(\ell, m)$ has the
same effect as~$R(\ell, m)$.

We define a circuit $C$ as follows:
\begin{enumerate}
\item Apply $R(1, k)$.
\item Apply $R'(k, n)$.
\end{enumerate}

First, note that the first half of $C$ sets $\bit{j} = a_{j+1}$ for $j
< k$, and $\bit{k} = a_1$.  Consequently, the second half of $C$
completes the rotation.  So, $C$ rotates $n$ wires as
desired.  Clearly the size of $C$ is $4n - 6$.

The only bit used both by $R(1, k)$ and $R'(k, n)$ is $\bit{k}$.  Note
that $R(1, k)$ is done accessing bit $\bit{k}$ after $k+3$
time-slices, and $R'(k, d)$ does not access $\bit{k}$ until time-slice
$n-k$.  Hence, the total depth of the circuit is $(k+3) + ((n-k) +
4) = n + 7$.

We can further reduce the depth to $n + 5$.  The last access of
$\bit{k}$ by $R(1, k)$ and the first access by $R'(k, n)$ both write
to $\bit{k}$.  These two operations commute with each other.  By
swapping the order, we can start $R'(k, n)$ two time-slices sooner.
\end{proof}

\section{Reversal}
\label{reverse-sec}

We now give a construction reversing the contents of $n$ wires
in depth $2n + 2$.  We then show that any such circuit has
depth at least $2n + 1$.

\subsection{Upper bound on reversal}
\label{reverse-upper-sec}

\begin{theorem}
\label{reverse-theorem}
We can reverse $n$ wires in depth $2n + 2$.  The
circuit has size $n^2 - 1$.
\end{theorem}

\begin{figure}[hb]
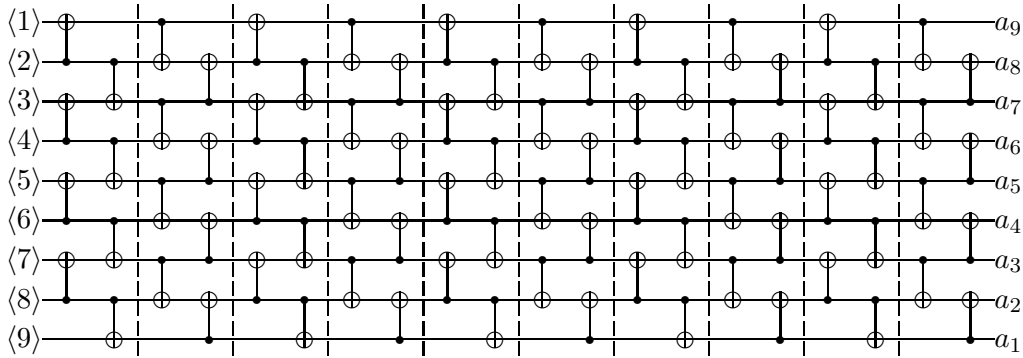

\begin{center}
\renewcommand{\arraystretch}{0}
\begin{tabular}{r@{}*{41}{c@{}}l}
\wire{$\bit{1}$} &\sn0&\xn2&\sm0&\sn0&\sm3&\cn2&\sm0&\sn0&\sm3&\xn2&\sm0&\sn0&\sm3&\cn2&\sm0&\sn0&\sm3&\xn2&\sm0&\sn0&\sm3&\cn2&\sm0&\sn0&\sm3&\xn2&\sm0&\sn0&\sm3&\cn2&\sm0&\sn0&\sm3&\xn2&\sm0&\sn0&\sm3&\cn2&\sm0&\sn0&\sn0& \wire{$a_9$} \\
\wire{$\bit{2}$} &\sn0&\cn1&\sm0&\cn2&\sm3&\xn1&\sm0&\xn2&\sm3&\cn1&\sm0&\cn2&\sm3&\xn1&\sm0&\xn2&\sm3&\cn1&\sm0&\cn2&\sm3&\xn1&\sm0&\xn2&\sm3&\cn1&\sm0&\cn2&\sm3&\xn1&\sm0&\xn2&\sm3&\cn1&\sm0&\cn2&\sm3&\xn1&\sm0&\xn2&\sn0& \wire{$a_8$} \\
\wire{$\bit{3}$} &\sn0&\xn2&\sm0&\xn1&\sm3&\cn2&\sm0&\cn1&\sm3&\xn2&\sm0&\xn1&\sm3&\cn2&\sm0&\cn1&\sm3&\xn2&\sm0&\xn1&\sm3&\cn2&\sm0&\cn1&\sm3&\xn2&\sm0&\xn1&\sm3&\cn2&\sm0&\cn1&\sm3&\xn2&\sm0&\xn1&\sm3&\cn2&\sm0&\cn1&\sn0& \wire{$a_7$} \\
\wire{$\bit{4}$} &\sn0&\cn1&\sm0&\cn2&\sm3&\xn1&\sm0&\xn2&\sm3&\cn1&\sm0&\cn2&\sm3&\xn1&\sm0&\xn2&\sm3&\cn1&\sm0&\cn2&\sm3&\xn1&\sm0&\xn2&\sm3&\cn1&\sm0&\cn2&\sm3&\xn1&\sm0&\xn2&\sm3&\cn1&\sm0&\cn2&\sm3&\xn1&\sm0&\xn2&\sn0& \wire{$a_6$} \\
\wire{$\bit{5}$} &\sn0&\xn2&\sm0&\xn1&\sm3&\cn2&\sm0&\cn1&\sm3&\xn2&\sm0&\xn1&\sm3&\cn2&\sm0&\cn1&\sm3&\xn2&\sm0&\xn1&\sm3&\cn2&\sm0&\cn1&\sm3&\xn2&\sm0&\xn1&\sm3&\cn2&\sm0&\cn1&\sm3&\xn2&\sm0&\xn1&\sm3&\cn2&\sm0&\cn1&\sn0& \wire{$a_5$} \\
\wire{$\bit{6}$} &\sn0&\cn1&\sm0&\cn2&\sm3&\xn1&\sm0&\xn2&\sm3&\cn1&\sm0&\cn2&\sm3&\xn1&\sm0&\xn2&\sm3&\cn1&\sm0&\cn2&\sm3&\xn1&\sm0&\xn2&\sm3&\cn1&\sm0&\cn2&\sm3&\xn1&\sm0&\xn2&\sm3&\cn1&\sm0&\cn2&\sm3&\xn1&\sm0&\xn2&\sn0& \wire{$a_4$} \\
\wire{$\bit{7}$} &\sn0&\xn2&\sm0&\xn1&\sm3&\cn2&\sm0&\cn1&\sm3&\xn2&\sm0&\xn1&\sm3&\cn2&\sm0&\cn1&\sm3&\xn2&\sm0&\xn1&\sm3&\cn2&\sm0&\cn1&\sm3&\xn2&\sm0&\xn1&\sm3&\cn2&\sm0&\cn1&\sm3&\xn2&\sm0&\xn1&\sm3&\cn2&\sm0&\cn1&\sn0& \wire{$a_3$} \\
\wire{$\bit{8}$} &\sn0&\cn1&\sm0&\cn2&\sm3&\xn1&\sm0&\xn2&\sm3&\cn1&\sm0&\cn2&\sm3&\xn1&\sm0&\xn2&\sm3&\cn1&\sm0&\cn2&\sm3&\xn1&\sm0&\xn2&\sm3&\cn1&\sm0&\cn2&\sm3&\xn1&\sm0&\xn2&\sm3&\cn1&\sm0&\cn2&\sm3&\xn1&\sm0&\xn2&\sn0& \wire{$a_2$} \\
\wire{$\bit{9}$} &\sn0&\sn0&\sm0&\xn1&\sm3&\sn0&\sm0&\cn1&\sm3&\sn0&\sm0&\xn1&\sm3&\sn0&\sm0&\cn1&\sm3&\sn0&\sm0&\xn1&\sm3&\sn0&\sm0&\cn1&\sm3&\sn0&\sm0&\xn1&\sm3&\sn0&\sm0&\cn1&\sm3&\sn0&\sm0&\xn1&\sm3&\sn0&\sm0&\cn1&\sn0& \wire{$a_1$} \\
\end{tabular}

\end{center}
\caption{Reversal of $9$ wires in depth $20$.}
\label{reverse-fig}
\end{figure}
An example of this construction for $n = 9$ appears in
Figure~\ref{reverse-fig}.

\begin{figure}
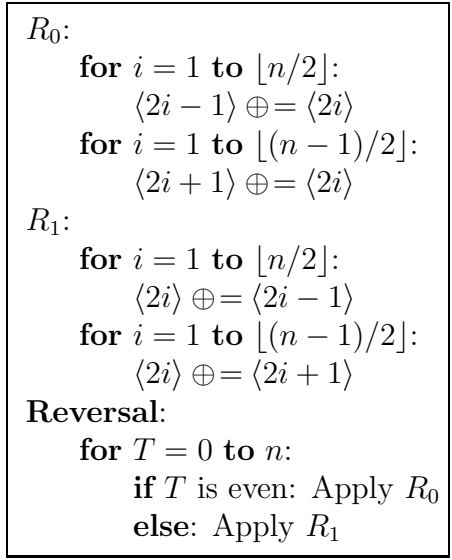

\centerline{\fbox{
\begin{minipage}{100in}
\begin{tabbing}
{\bf for} \= {\bf for} \= {\bf for} \= \kill
$R_0$: \+ \\
{\bf for} $i = 1$ {\bf to} $\floor{n/2}$: \\ 
\> $\bit{2i - 1} \xoreq \bit{2i}$ \\
{\bf for} $i = 1$ {\bf to} $\floor{(n-1)/2}$: \\
\> $\bit{2i + 1} \xoreq \bit{2i}$ \- \\
$R_1$: \+ \\
{\bf for} $i = 1$ {\bf to} $\floor{n/2}$: \\ 
\> $\bit{2i} \xoreq \bit{2i - 1}$ \\
{\bf for} $i = 1$ {\bf to} $\floor{(n-1)/2}$: \\
\> $\bit{2i} \xoreq \bit{2i + 1}$ \- \\
{\bf Reversal}: \+ \\
{\bf for} $T = 0$ {\bf to} $n$: \\
\> {\bf if} $T$ is even:  Apply $R_0$ \\
\> {\bf else}:  Apply $R_1$
\end{tabbing}
\end{minipage}}}
\caption{Reversal circuit.  For $n > 2$,
the subcircuits $R_0$ and $R_1$ each have depth 2.}
\label{reverse-code-fig}
\end{figure}

\begin{proof}
The reversal circuit is described in Figure~\ref{reverse-code-fig}.
The subcircuit $R_0$ adds the contents of each wire with an even index
into its neighbors; the subcircuit $R_1$ adds the odd-indexed wires
into their neighbors.  We alternate between these two operations.

For a given value of $i$, we will keep track of which wires depend on
$a_i$ over time.  First suppose that $i$ is even.  To simplify matters,
we will see what the effect of successive applications of $R_0$ and
$R_1$ would be if there were wires corresponding to arbitrarily small
and large integers.  After the first application of $R_0$, since $i$
is even, we perform $\bit{i-1} \xoreq \bit{i}$ and $\bit{i+1} \xoreq
\bit{i}$, so $a_i$ gets added to $\bit{i-1}$ and $\bit{i+1}$.  Thus
$a_i$ affects $\bit{i-1}$, $\bit{i}$, and $\bit{i+1}$.  After we next
apply $R_1$, $\bit{i-1}$ and $\bit{i+1}$ are added to their
neighboring wires, so $a_i$ affects $\bit{i-2}$ through $\bit{i+2}$.
(The effects of the two additions to $\bit{i}$ cancel.)  In general,
after applying $R_0$ and $R_1$ a total of $t$ times,
$a_i$ will affect $\bit{i-t}$ through $\bit{i+t}$.

Now let us take into account the fact that we only have wires
$\bit{1}$ through $\bit{n}$.  During the $i$th application of an
$R$-subcircuit, we cannot add $\bit{1}$ to $\bit{0}$, since the
latter does not exist, so $\bit{1}$ is still the lowest-numbered wire
affected by $a_i$.  During the $(i+1)$th application of an
$R$-subcircuit, $\bit{2}$ is added to $\bit{1}$, so $\bit{1}$ no
longer depends on $a_i$, and $\bit{2}$ is now the first wire affected
by $a_i$.  Therefore, after $t$ applications of $R$-subcircuits, for $t
\ge i$, the lowest-numbered wire affected by $a_i$ is $\bit{t-i+1}$.
Similarly, for $t > n-i$, the highest-numbered wire affected by $a_i$
is $\bit{n - (t-1-(n-i))} = \bit{2n-t-i+1}$. (We can see this by
interchanging $i$ and $n+1-i$, relabeling the wires in the opposite
order, and interchanging $R_0$ and $R_1$ if $n$ is even.)  That is, for
$t$ bigger than both $i$ and $n-i$ (and not too large), $a_i$ will
affect exactly the wires $\bit{t-i+1}$ through $\bit{2n-t-i+1}$ after
$t$ applications of $R$-subcircuits.

Our circuit applies $R$-subcircuits a total of $n+1$ times.  After
$n$ of these, $a_i$ affects exactly wires $\bit{n+1-i}$ through
$\bit{n+1-i}$; that is, $a_i$ affects only $\bit{n+1-i}$.  Since this
$n$th application writes to wires of the opposite parity of $\bit{n+1-i}$, the
$(n+1)$th application will write to wires of the same parity as
$\bit{n+1-i}$, and $\bit{n+1-i}$ will still be the only wire affected
by~$a_i$.

Finally, we consider the case with $i$ odd. After the first
application of an $R$-subcircuit, $\bit{i}$ is still the only wire
affected by $a_i$. Then, as above, after $n$ more applications,
$\bit{n+1-i}$ is the sole wire affected by~$a_i$.

We have shown that, after our circuit runs, the wire $\bit{n+1-i}$
will depend on $a_i$, but no other wire $\bit{n+1-j}$ for $j \ne i$
will.  Turning this around, we see that the final value of $\bit{n+1-i}$
does not depend on $a_j$ for $j \ne i$, so that this final value must,
in fact, be equal to $a_i$. We have performed reversal, as desired.
\end{proof}

For $n = 2$, the subcircuits $R_0$ and $R_1$ each have depth $1$, so the
overall depth of our reversal circuit is $3$.  For $n > 2$, the
depth is $2n + 2$.

\subsection{Lower bound on reversal}
\label{reverse-lower-sec}

For $2 \le n \le 6$, computer searches confirm that the above
construction is optimal.  We conjecture that the depth of any circuit
performing reversal for $n \ge 3$ is at least $2n + 2$.  We now show that
any such circuit has depth at least $2n + 1$.

\begin{lemma}
\label{reverse-lower-lemma}
For any $k \le n/2$, any circuit reversing $n$ wires contains at least
$2k+1$ gates between wires $\bit{k}$ and $\bit{k+1}$
and also between wires $\bit{n-k}$ and $\bit{n - k + 1}$.
If $k$ is not $n/2$, then there must be at least $2k+1$ such gates before
the last time-slice.
\end{lemma}

\begin{proof}
Let $R$ be a circuit reversing $\bit{1}, \dots, \bit{n}$.
We show that $R$ must have at least $2k+1$ gates between wires $\bit{k}$
and $\bit{k+1}$; the proof for $\bit{n-k}$ and $\bit{n-k+1}$ is analogous.

We write the contents of the wires at any given time as a block matrix
\begin{equation}
\label{k-block-eq}
M = \begin{pmatrix} W & X \\ Y & Z \end{pmatrix},
\end{equation}
where $W$ is $k \times k$, $X$ is $k \times (n-k)$, $Y$ is $(n-k) \times k$,
and $Z$ is $(n-k) \times (n-k)$.  The matrix $M$ changes as we apply $R$.
Initially, $W$ and $Z$ are identity matrices of sizes $k$ and $n-k$, and
$X$ and $Y$ are $0$.  When we conclude, $W$, $X$, $Y$, and $Z$ have
ranks $0$, $k$, $k$, and $n-2k$, respectively.

The ranks of $W$, $X$, $Y$, and $Z$ are affected only by gates between
wires $\bit{k}$ and $\bit{k+1}$.  Each upward gate $\bit{k} \xoreq \bit{k+1}$
changes the ranks of $W$ and $Y$ by at most $1$, and each downward
gate $\bit{k+1} \xoreq \bit{k}$ changes the ranks of $X$ and $Z$ by at most
$1$.  Each of the four ranks has to change by $k$.  We conclude that
there are at least $k$ upward and $k$ downward gates in $R$.

Furthermore, suppose that the first gate between $\bit{k}$ and $\bit{k+1}$
is upward.  At this point $X$ is still $0$, so the gate cannot
affect the rank of $W$; the circuit $R$ requires
$k$ more upward gates.  Similarly, if the first gate is downward,
it cannot affect the rank of $Z$, and $R$ requires $k$
additional downward gates.  Hence, there must be at least $2k+1$ gates between
$\bit{k}$ and $\bit{k+1}$.

Finally, if $k$ is not exactly $n/2$, then any gate
in the last time-slice cannot affect any of the ranks of $W, X, Y, Z$,
so all of the gates accounted for above must occur in earlier time-slices.
\end{proof}

\begin{theorem}
\label{reverse-lower-thm}
Reversing $n \ge 3$ wires requires depth at least $2n + 1$ and size
at least $\floor{\frac12 n^2} + n$.
\end{theorem}

\begin{proof}
First, suppose $n = 2r$.  Given any circuit $R$ for reversal,
we obtain another reversal circuit by vertically flipping the last
time-slice of $R$ (that is, conjugating it by reversal) and moving it to
the beginning of the circuit.  We may therefore assume, without loss
of generality, that the last time-slice contains a gate between
$\bit{r+1}$ and $\bit{r+2}$.

By Lemma~\ref{reverse-lower-lemma},
there are at least $2r+1$ gates between wires $\bit{r}$
and $\bit{r+1}$ and at least $2(r-1) + 1 = 2r-1$ gates between wires $\bit{r+1}$
and $\bit{r+2}$ before the last time-slice.
Hence, there are at least $4r + 1 = 2n + 1$ gates
involving $\bit{r+1}$, giving the lower bound on depth.  If we sum over
all locations, we find that the total number of gates is at least
\[
2r + 1 + 2\sum_{i=1}^{r-1}(2i+1) + 1 = \frac{n^2 + 2n}{2}.
\]

Second, suppose $n = 2r+1$.  Again, we may assume that the last time-slice
contains a gate between $\bit{r+1}$ and $\bit{r+2}$.
Now we have at least $2r+1$ gates between wires $\bit{r}$ and $\bit{r+1}$
and at least $2r+2$ gates between $\bit{r+1}$ and $\bit{r+2}$.  This gives
a total of $4r+3 = 2n+1$ gates involving $\bit{r+1}$, meaning we must have
at least $2n+1$ time-slices.  The total number of gates is at least
\[
2\sum_{i=1}^r(2i+1) + 1 = \frac{n^2 + 2n - 1}{2}.
\]
\end{proof}


\section{Permutation}
\label{permute-sec}

We now discuss the more general problem of permuting the $n$ input
bits.  It is easier to visualize the problem by imagining that the
wire $\bit{i}$ contains the data $a_i$ with the attached label $\sigma(i)$.
We then wish to sort the data by their labels.  When we finish, the wire
$\bit{i}$ will have the label $i$, and hence the bit $a_{\sigma^{-1}(i)}$,
as desired.

\begin{theorem}
\label{permute-thm}
For any $\sigma \in S_n$, there is a circuit implementing
$\sigma$ with depth at most $3n$ and size at most $3\binom{n}{2}$.
\end{theorem}

\begin{proof}
It is convenient to pretend that our basic operation is a swap of
two adjacent bits; we can implement such a swap using three of
our standard gates.  Initially, our labels are in the order
$\sigma(1), \dots, \sigma(n)$; after each swap, the order changes.
When the circuit completes, we want the labels to be sorted.

To effect the swaps, we use an $n$-bit {\em sorting network\/}.
The basic gate is a {\em conditional
swap\/} on $\bit{i}$ and $\bit{j}$:\ if $i < j$ but the label on
$\bit{i}$ is larger than the label on $\bit{j}$, then we swap the
contents and labels of the two wires.  A network of conditional swaps
is a sorting network if, for any (valid) initial assignment of labels,
at the end wire $\bit{i}$ has label $i$.  We are interested in sorting
networks using only conditional swaps on adjacent wires.
See~\cite[Section 5.3.4]{knuth-volume-3-2nd} for more discussion.

Suppose we have a $n$-bit sorting network of depth $d$ and size
$s$, in which each conditional swap is between two adjacent wires.  We
will perform each swap only if the label of the second bit is less
than that of the first bit.  Since we know $\sigma$ in advance, we
know which swaps to leave in the network and which to leave out.  The
result will be a swap network with depth at most $d$ and size at most
$s$.  The corresponding circuit has depth at most $3d$ and size at
most~$3s$.

\begin{figure}[h]
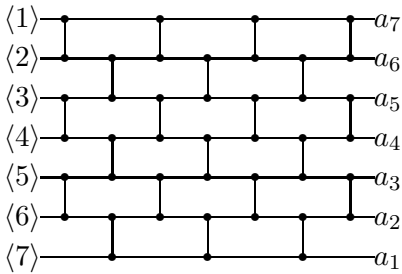

\begin{center}
\renewcommand{\arraystretch}{0}
\begin{tabular}{r@{}*{15}{c@{}}l}
\wire{$\bit{1}$} &\sn0&\cn2&\sm0&\sn0&\sm0&\cn2&\sm0&\sn0&\sm0&\cn2&\sm0&\sn0&\sm0&\cn2&\sn0& \wire{$a_7$} \\
\wire{$\bit{2}$} &\sn0&\cn1&\sm0&\cn2&\sm0&\cn1&\sm0&\cn2&\sm0&\cn1&\sm0&\cn2&\sm0&\cn1&\sn0& \wire{$a_6$} \\
\wire{$\bit{3}$} &\sn0&\cn2&\sm0&\cn1&\sm0&\cn2&\sm0&\cn1&\sm0&\cn2&\sm0&\cn1&\sm0&\cn2&\sn0& \wire{$a_5$} \\
\wire{$\bit{4}$} &\sn0&\cn1&\sm0&\cn2&\sm0&\cn1&\sm0&\cn2&\sm0&\cn1&\sm0&\cn2&\sm0&\cn1&\sn0& \wire{$a_4$} \\
\wire{$\bit{5}$} &\sn0&\cn2&\sm0&\cn1&\sm0&\cn2&\sm0&\cn1&\sm0&\cn2&\sm0&\cn1&\sm0&\cn2&\sn0& \wire{$a_3$} \\
\wire{$\bit{6}$} &\sn0&\cn1&\sm0&\cn2&\sm0&\cn1&\sm0&\cn2&\sm0&\cn1&\sm0&\cn2&\sm0&\cn1&\sn0& \wire{$a_2$} \\
\wire{$\bit{7}$} &\sn0&\sn0&\sm0&\cn1&\sm0&\sn0&\sm0&\cn1&\sm0&\sn0&\sm0&\cn1&\sm0&\sn0&\sn0& \wire{$a_1$} \\
\end{tabular}

\end{center}
\caption{$7$-wire sorting network in depth $7$.}
\label{sort-fig}
\end{figure}


It merely remains to construct an efficient sorting network using only
conditional swaps of adjacent wires.  We use the odd--even transposition
sort.\footnote{See Knuth~\cite[Exercise 5.3.4.37]{knuth-volume-3-2nd}
for a proof of correctness and a brief history.}  It has $n$ steps,
alternating between performing all conditional swaps of the form $(2j-1, 2j)$
and performing all conditional swaps of the form $(2j, 2j + 1)$. 
An example with $n = 7$ is depicted in Figure~\ref{sort-fig}.  We have
$s = n(n-1)/2$ and $d = n$ (unless $n = 2$, in which case~$d=1$).
\end{proof}

We observe that the above sorting network achieves the optimal $d$ and
$s$.  First, note that each swap reduces the number of inversions by
at most one.  Since $\sigma$ can have up to $\binom{n}{2}$ inversions, we
must have $s \ge \binom{n}{2}$.

In addition, in an optimal sorting network, we cannot perform the same
swap in consecutive time-slices.  Thus, in any two consecutive
time-slices, we can perform at most $n-1$ swaps.  Hence, for all $n >
2$, we need at least $\floor{n/2}$ pairs of time-slices to
accommodate $(n-1)\floor{n/2}$ gates, plus (at least) one more
time-slice if $n$ is odd.  Hence, for all $n > 2$ we have $d \ge n$.

Clearly, for a particular permutation, we may be able to do better
than Theorem~\ref{permute-thm} would suggest; see, for example,
Sections~\ref{swap-sec}, \ref{rotate-sec}, and~\ref{reverse-sec}.  A
more difficult problem is determining the minimum depth for the worst
possible~$\sigma$.

For $n \le 6$, reversal is at least as hard as any other permutation:\ 
we can implement any permutation in depth $2n + 2$.  We do not know
whether this pattern holds for larger~$n$.

\suppressfloats

\section{Arbitrary Matrices}
\label{gln-sec}

As noted in the Introduction, any circuit on $n$ wires
made up of {\cnot} gates computes a matrix in $\GL_n(2)$.  Conversely,
given a matrix, it is straightforward to build a circuit with depth
$O(n^2)$.

More concretely, we suppose the initial state of the wires is
described by the identity matrix $I$; 
each wire $\bit{i}$ contains the basis vector $\e_i$.
If a circuit $C$ applied to this initial state $I$ results in
state $M$, we say that $C$ performs
the transformation $M$.  This map from circuits to matrices is a
homomorphism.  

The problems of building a circuit performing $M$ and a circuit
performing $M^{-1}$, for an arbitrary invertible matrix $M$, are
equivalent.  Notationally, we find the latter more convenient.
Instead of building a circuit to perform $M$, we suppose the wires
start in state $M$, and we construct a circuit to ``undo'' $M$ and
restore $I$.  The reverse of this circuit will perform $M$.

In this section we give a constructive proof of the following result:

\begin{theorem}\label{main-thm}
Let $M$ be a matrix in $\GL_n(2)$.  Then
there is a circuit computing $M$ with depth at most $5n$.
\end{theorem}


Our construction uses the concept of a ``northwest''-triangular matrix.

\begin{definition}
An $n \times n$ matrix $M$ 
is {\em northwest-triangular\/} if $M_{ij} = 0$ for all $i + j > n+1$.
\end{definition}

We discuss the building blocks of our circuit in Section~\ref{boxes-sec}.
In Sections~\ref{sec-5n} and~\ref{sec-6n}, we prove the following
propositions:

\begin{proposition}\label{clearnet}
Let $M$ be in $\GL_n(2)$.  Given an $n$-wire sorting network of depth $d$,
we can construct a circuit $C$ of depth $2d$ such that $MC$ is
northwest-triangular.
\end{proposition}

\begin{proposition}\label{revnet}
Let $N$ be an invertible northwest-triangular matrix.  Given an $n$-wire
sorting network of depth $d$, we can construct
a circuit $R$ of depth $3d$ with $NR = I$.
\end{proposition}

\begin{proof}[Proof of Theorem~\ref{main-thm}]
Let $M$ be any matrix in $\GL_n(2)$.  By Proposition~\ref{clearnet},
using the odd-even transposition network of depth $n$, there
is a circuit $C$ of depth $2n$ such that $MC$ is northwest-triangular.
By Proposition~\ref{revnet} (using the same network),
there is a circuit $R$ of depth $3n$ with $MCR = I$.  Then $R^{-1}C^{-1}$
computes $M$.
\end{proof}


The maximum possible size (that is, number of gates) of a depth-$d$
circuit is $d \floor{n/2}$.  The {\em density\/} of a circuit is its
size divided by this maximum.  
The construction of Theorem~\ref{main-thm}
has size about ${\frac52}n^2$ and density $1$.  
We also have a construction with size about $2n^2$ and density $1/2$.
(See Section~\ref{sec-4n}.)
Note that, if we could construct a circuit with size $2n^2$ and
density $1$, we would have a solution with depth $4n$.  We discuss
this, and other reasons why we conjecture that circuits of depth
$4n + O(1)$ may be possible, in Section~\ref{sec-4n}.

\subsection{Boxes}
\label{boxes-sec}


The building blocks for our circuits will be not individual
{\cnot} gates, but {\em boxes\/}:

\begin{definition}
A {\em box\/} is a subcircuit on two adjacent wires $\bit{i}$
and $\bit{i+1}$.
\end{definition}

Every box performs some operation in $\GL_2(2)$.
If $u$ and $v$ are the contents of the two input wires to a box, then
the two output wires contain distinct elements of $\{u, v, u\xor v\}$.
Some researchers (for example,~\cite{FDH}) compute the costs of quantum
circuits by counting arbitrary $2$-qubit interactions; in such a model,
the box is the fundamental unit.

\begin{table}[ht!]
\caption{Depth of implementing boxes with input $u, v$.}
\label{box-table}
\begin{center}
\begin{tabular}{cc|c}
First & Second & \\
Output & Output & \raiseintable{Depth} \\ \hline
$u$ & $v$ & $0$ \\
$u$ & $u \xor v$ & $1$ \\
$u \xor v$ & $v$ & $1$ \\
$u \xor v$ & $u$ & $2$ \\
$v$ & $u \xor v$ & $2$ \\
$v$ & $u$ & $3$
\end{tabular}
\end{center}
\end{table}

The depth of a box depends on the two output vectors, as shown in
Table~\ref{box-table}.  If we want to perform an arbitrary operation
in $\GL_2(2)$, then the depth of our box could be as large as~$3$.
However, if we only specify one of the two outputs, and allow the
other output to take whichever value is more convenient, we see
that we can make do with boxes of depth $2$.

\subsection{Clearing Networks}
\label{sec-5n}

We now prove Proposition~\ref{clearnet}.  We use a sorting network
to build a system of depth-2 boxes
to convert any matrix into northwest-triangular form.
 
\begin{proof}[Proof of Proposition~\ref{clearnet}]
We first perform a lower-triangular
basis change; this does not involve changing
the contents of any wires, but merely describes them differently.
We then construct a circuit.

Let $V = \ff^n$ be the space containing our wires.
We define a {\em lexicographic order\/} on $V$.  For $u, v \in V$,
we write $u \prec v$ if there exists $k$ such that $u \cdot \e_k = 0$,
$v \cdot \e_k = 1$, and, for all $j > k$, 
$u \cdot \e_j = v \cdot \e_j$.

For each $i$, let $v_i$ be the lexicographically least element of
$a_i \xor \Span\{a_j : i < j \le n\}$.  Note that this is a
lower-triangular basis change:\ for each $i$,
$a_i \in v_i \xor \Span\{v_j : i < j \le n\}$.

For each $i$, let $\pi(i)$ be the smallest $j$ such that
$v_i \cdot \e_{n+1-j} = 1$.  By construction, $\pi$ is a permutation:\
for any $k < \ell$, we must have $v_k \prec v_k \xor v_\ell$, and
therefore $\pi(k) \ne \pi(\ell)$.

Let $w_j = v_{\pi^{-1}(j)}$.  The $w_j$ satisfy
\[
w_j \in \e_{n+1-j} \xor \Span\{\e_k : 1 \le k < n+1-j\}.
\]
Attach to each wire $\bit{i}$ the label $\pi(i)$. 
We maintain the following invariant:
\begin{itemize}
\item If a wire has value $\sum \alpha_j w_j$,
and $k$ is the label on some lower-numbered wire, then $\alpha_k = 0$.
\end{itemize}

The invariant is true initially because $\sum\alpha_j
w_j=\sum\alpha_{\pi(i)}v_i$ and the basis change is lower-triangular.
If we sort the labels and maintain the invariant,
then when we are done, the value of wire $\bit{i}$ is in
\[
w_i \xor \Span\{w_j : i < j \leq n \}
= \e_{n+1-i} \xor \Span\{\e_j : 1 \le j < n + 1 - i\},
\]
so the wires specify a northwest-triangular matrix.

We now build a circuit $C$ that sorts the labels
while maintaining the invariant.  We start with a sorting network $S$
of depth $d$ and replace each conditional swap
in $S$ by a box.  Suppose we have two inputs to a box:\ $\bit{i}$
has value $u$ and label $j$, and $\bit{i+1}$ has value $v$ and label $k$.
If $j < k$, we do nothing.  If $j > k$, we swap the two labels,
and we also perform a box as described below.
When the network concludes, we will have sorted the labels, as desired.

Let $W$ be the span of all $w_\ell$ for $\ell \ne k$.  The space $W$ has
codimension $1$, so at least one of $\{u, v, u \xor v\}$ lies in $W$.
We can perform a box on $\bit{i}$ and $\bit{i+1}$ that writes a vector in $W$
to wire $\bit{i+1}$.
This maintains the invariant for wires
$\bit{i}$ and $\bit{i+1}$, as desired, and other wires are unaffected.

Each box in $C$ comes from a conditional swap in $S$.  We are specifying
only one output of each box, so each box has depth at most $2$.
Hence, the depth of $C$ is at most $2d$.
\end{proof}

\subsection{Reversal Networks}
\label{sec-6n}


We now prove Proposition~\ref{revnet}:\ we reduce any
northwest-triangular matrix to the identity.  As before, we use a
sorting network to build a system of boxes.  However, in this
case our boxes are permitted to have depth $3$.

\begin{proof}[Proof of Proposition~\ref{revnet}]
We first label each input wire $\bit{i}$ with $n+1-i$.
We take a sorting network $S$ of depth $d$ and convert $S$
to a reversal network; we include exactly
those conditional swaps that are used when input wire $\bit{i}$ has
label $n+1-i$.  (The new network will have size $\binom{n}{2}$;
if $S$ has the minimal size $\binom{n}{2}$,
then it already is a reversal network.)  We make each
remaining swap unconditional:\ we definitely swap the two labels.

Consider a swap between $\bit{i}$, with value $u$ and label $k$,
and $\bit{i+1}$, with value $v$ and label $j$.  Note that $k > j$.
If $u \cdot \e_j = 0$, we replace the swap with a
depth-$3$ box exchanging $u$ and $v$.
If $u \cdot \e_j = 1$, then we replace the swap with
the depth-$2$ box that
first adds $u$ into $v$ and then adds $u \xor v$
into $u$; this has the effect of replacing $u$ by $u\oplus v$ and then
exchanging (the new) $u$ and $v$.

We claim that this circuit maintains the following invariants:
\begin{enumerate}
\item If $u$ is on the wire with label $k$, then
$u \cdot \e_k = 1$ and $u \cdot \e_\ell = 0$ for $\ell > k$.
\item If $u$ is on $\bit{i}$ with label $k$, and
$\bit{h}$ has label $j$, with $h < i$ and $j < k$, 
then $u \cdot \e_j = 0$.
\end{enumerate}

Initially, $\bit{i}$ has label $n+1-i$.  The first
invariant holds because $N$ is an invertible northwest-triangular matrix.
The second
invariant holds vacuously, as there are no such pairs of wires. 

What is the effect of a single box between wires $\bit{i}$ and
$\bit{i+1}$ with values $u$ and $v$ and labels $k$ and $j$?
The box necessarily maintains
the first invariant.  Swapping $u$ and $v$ has no effect.  The step
replacing $u$ by $u \xor v$ also is not a problem: $k > j$ implies $v
\cdot \e_\ell = 0$ for all $\ell \geq k$.

This circuit also maintains the second invariant.  It clearly still
holds for all wires besides $\bit{i}$ and $\bit{i+1}$.
The value $v$ and label $j$ move unchanged from wire $\bit{i+1}$ to
wire $\bit{i}$, so it holds for $\bit{i}$ as well.
If label $\ell$ is on wire $\bit{h}$, with $h < i$ and $\ell < k$, then 
$u \cdot \e_\ell = 0$.  Also, $v \cdot \e_\ell = 0$, either
by the first invariant if $\ell > j$, or by the second 
if $\ell < j$, so $(u \xor v) \cdot \e_\ell = 0$.  Finally,
we have designed the box so that the output value on $\bit{i+1}$,
either $u$ or $u \xor v$, is orthogonal to $\e_j$.

When $R$ concludes, the labels are in order; wire $\bit{i}$
has label $i$.
The two invariants then imply that $\bit{i}$ contains $\e_i$; that
is, we have reached the identity matrix.
\end{proof}

\subsection{Lower Bounds}
\label{general-lower-sec}

By Theorem~\ref{reverse-lower-thm}, reversal requires depth $2n + 1$.  Hence,
we have already shown that the minimum depth for the worst-case
matrix in $\GL_n(2)$ is at least $2n + 1$.  We now argue that almost all invertible
matrices require about this depth.  By ``almost all invertible matrices''
we mean
a proportion of elements of $\GL_n(2)$ tending to $1$ as $n$ goes to $\infty$.
First we quote a well-known result on ranks of random matrices.

\begin{theorem}
As $n$ goes to $\infty$, the proportion of $n \times n$ matrices over $\ff$
having rank at most $n-c$ is $O(2^{-c^2})$.
\end{theorem}

\begin{proof}[Sketch of proof]
This follows from the fact that the number of $n \times n$ matrices of rank $k$
is equal to the square of the number of $n \times k$ matrices of rank $k$
divided by the number of invertible $k \times k$ matrices.
To count these numbers of matrices, we use a standard formula
of Landsberg~\cite{Land}; see Stanley~\cite[Section 1.3]{Stan} for a
more recent exposition.
\end{proof}

\begin{lemma}
Let $\epsilon > 0$ be given.  For almost all matrices $M$ in $\GL_n(2)$,
every circuit implementing $M$ has, for each integer $k$ with $1 \le k \le n/2$,
at least $2k - \epsilon n$ gates between
wires $\bit{k}$ and $\bit{k+1}$ and also between wires $\bit{n-k}$ and $\bit{n-k+1}$.
\end{lemma}


\begin{proof}
The proof uses the same technique as that of Lemma~\ref{reverse-lower-lemma}.
As before, we consider wires $\bit{k}$ and $\bit{k+1}$.
Choose $M \in \GL_n(2)$ uniformly at random,
and consider a circuit implementing $M$. 
We write the contents of the wires at any time as a block matrix, as
in~\eqref{k-block-eq}.  Initially, $X$ and $Y$ are $0$, and at the conclusion
of the circuit, $X$ and $Y$ are two blocks of our matrix $M$.  For large enough
values of $n$ and for a random choice of $M$,
we expect $X$ and $Y$ each to have rank at least $k - (\epsilon/2)n$;
since each gate between $\bit{k}$ and $\bit{k+1}$ changes the total rank
of $X$ and $Y$ by at most $1$, we have at least $2k - \epsilon n$ such gates.
\end{proof}

Counting gates between different pairs of bits yields the following theorem:

\begin{theorem}\label{general-lower-thm}
Let $\epsilon > 0$ be given.  For almost all matrices $M$ in $\GL_n(2)$
every circuit implementing $M$ requires depth at least $(2 - \epsilon)n$
and size at least $(1 - \epsilon)n^2$.
\end{theorem}

A more careful analysis shows that the proportion of
matrices in $\GL_n(2)$ that can be implemented in depth at most $2n - m$
is $O(2^{-(m-2)^2/8})$.

\section{Open Questions}
\label{conclude-sec}
\label{sec-4n}

Let the ``depth'' of a matrix be the minimum depth of any circuit implementing
the matrix.  We have shown that the maximum depth of a matrix in $\GL_n(2)$
lies between $2n + 1$ and $5n$.  A natural question is
whether we can close this gap.

For several reasons, the authors feel that the maximum depth may be
only $4n + O(1)$.  First, we consider circuit size:  By Theorem~\ref{reverse-lower-thm},
reversal requires at least $n^2/2$ gates, and the construction of
Section~\ref{gln-sec} computes any matrix in at most $5n^2/2$ gates.
Bob Beals~\cite{beals} has shown that one can compute any matrix
in only $2n^2$ gates.  If we could pack these gates into a rectangular
array, we could implement the matrix in depth $4n$.

More precisely, let $\nabla$ be the set of all matrices implementable
as $\nabla$-shaped arrays of $\binom{n}{2}$ depth-$2$ boxes.  A circuit
in $\nabla$ has size at most $n^2$ and depth at most $4n$.
Beals showed~\cite{beals} that $\nabla^2 = \GL_n(2)$:\ given $M$,
he builds two circuits, one on either side of $M$, so that the product is the
identity. Let $\Xi$ be the set of all rectangular arrays of $\binom{n}{2}$
depth-$2$ boxes; a circuit in $\Xi$ has size at most $n^2$ and depth
at most $2n$.  If we could similarly construct circuits in $\Xi$ on
either side of a matrix $M$ to reduce it to the identity, then $\Xi^2$ would
equal $\GL_n(2)$, and we could implement any matrix in depth $4n$.

By Proposition~\ref{clearnet}, we can use $\Xi$ to reduce any matrix
to northwest-triangular form.  It is interesting to note that the subgroup of
upper (or lower) triangular matrices in $\GL_n(2)$
has index $\prod_{i=1}^n (2^i - 1)$,
but its order is only $\prod_{i=1}^n 2^{i-1} = 2^{(n^2 - n)/2}$.
Thus, one could argue that we are ``working harder'' to reduce a general
matrix to northwest triangular form (in depth $2n$) than to reduce the
triangular matrix to the identity (in depth $3n$).  One could
imagine that the latter reduction should be possible in
the same depth as the former, providing further
evidence that $\Xi^2$ might contain all of $\GL_n(2)$.

We performed exhaustive computer experiments for $n$ up to $6$.
The maximum depths are shown in Table~\ref{computer-table}.
While we are reluctant to draw inferences from such limited data, these values
suggest that the maximum depth may be as small as $2n + O(1)$.
In other words, the lower bound of Theorems~\ref{reverse-lower-thm}
and~\ref{general-lower-thm} may be tight up to an additive constant.

\begin{table}[ht!]
\caption{Maximum depth of a matrix in $\GL_n(2)$ for
$n \le 6$ obtained by exhaustive search.}
\label{computer-table}
\begin{center}
\begin{tabular}{c|ccccc}
$n$ & 2 & 3 & 4 & 5 & 6 \\ \hline
depth & 3 & 8 & 10 & 13 & 14
\end{tabular}
\end{center}
\end{table}

We also give some additional open questions:
\begin{itemize}
\item For addition, swap, and rotation, we have an upper bound
for depth of $n + O(1)$ and a lower bound of $n-1$.  What is the
correct additive constant?
\item For $n \ge 3$, the optimal depth for reversal is either $2n+1$
or $2n+2$.  Which is correct?
\item What is the correct depth for a general permutation?  For
small $n$, reversal is at least as hard as any other permutation;
does this hold for all~$n$?
\item For general matrices, we have a lower bound on size of
$n^2/2$ and an upper bound of $2n^2$~\cite{beals}.  What is the
correct answer?
\item As noted earlier,
some researchers~\cite{FDH,kutin} use the box as the basic unit
of computation; in this model, we have a lower bound on depth
of $n+1$ (for reversal) and an upper bound of $2n$.  What is the
correct answer? 
\item Are there other classes of operations that can be
implemented efficiently in this model?
\end{itemize}

The last question above is perhaps the most intriguing.  Our focus was
on selecting natural operations on $n$ wires and then determining
their depth.  An alternative approach would be to consider all
circuits of a given depth and see what other useful operations can be
performed.  Such an analysis might suggest new efficient circuits for
arbitrary matrices and might even yield new approaches to quantum
circuit design.

\section*{Acknowledgments}

Tom Draper helped with our early work on addition, swap, and rotation.
Bob Beals made many helpful suggestions and pointed out the generalization
from Theorem~\ref{reverse-lower-thm} to Theorem~\ref{general-lower-thm}.

\bibliography{distance-d}
\bibliographystyle{plain}

\end{document}